\documentclass[aps,pra,twocolumn,showpacs,preprintnumbers,amsmath,amssymb,footinbib]{revtex4}
\usepackage{mathtools}

\DeclareMathAlphabet{\mathcalligra}{T1}{calligra}{m}{n}
\DeclareMathAlphabet{\mathpzc}{OT1}{pzc}{m}{it}

\usepackage{bbold}
\usepackage{graphicx,epsfig}
\usepackage{bm}
\usepackage{dcolumn}
\usepackage{xcolor}
\usepackage[breaklinks=true,colorlinks,citecolor=blue,linkcolor=blue,urlcolor=blue]{hyperref}

\def\nat#1#2#3{Nature {\bf #1}, #2 (#3)}

\def\sc#1#2#3{Science {\bf #1}, #2 (#3)}

\def\rmp#1#2#3{Rev. Mod. Phys. {\bf #1}, #2 (#3)}
\def\prl#1#2#3{Phys. Rev. Lett. {\bf #1}, #2 (#3)}
\def\pra#1#2#3{Phys. Rev. A {\bf #1}, #2 (#3)}

\def\epjd#1#2#3{Eur. Phys. J. D {\bf #1}, #2 (#3)}

\def\pla#1#2#3{Phys. Lett. A {\bf #1}, #2 (#3)}

\def\ejp#1#2#3{Eur. J. Phys. {\bf #1}, #2 (#3)}

\def\zp#1#2#3{Z. Phys. {\bf #1}, #2 (#3)}

\def\noi{\noindent}
\def\bc{\begin{center}}
\def\ec{\end{center}}
\newcommand{\bea}{\begin{equation}}
\newcommand{\eea}{\end{equation}\noi}
\newcommand{\ber}{\begin{eqnarray}}
\newcommand{\eer}{\end{eqnarray}\noi}
\begin{document}
\title{Artificial magnetism for a harmonically trapped Fermi gas in a synthetic magnetic field}

\author{Shyamal Biswas}\email{sbsp [at] uohyd.ac.in}
\author{Avijit Ghosh}
\author{Soumyadeep Majumder}
\affiliation{School of Physics, University of Hyderabad, C.R. Rao Road, Gachibowli, Hyderabad-500046, India}

\date{\today}

\begin{abstract}
We have analytically explored the artificial magnetism for a 3-D spin-polarized harmonically trapped ideal Fermi gas of electrically neutral particles exposed to a uniform synthetic magnetic field. Though polarization of the spin is necessary for trapping electrically neutral atoms in a magneto-optical trap, Pauli paramagnetism can not be studied for the spin-polarized Fermi system. However, it is possible to study Landau diamagnetism and de Haas-van Alphen effect for such a system. We have unified the artificial Landau diamagnetism and the artificial de Haas-van Alphen effect in a single framework for all temperatures as well as for all possible magnitudes of the synthetic magnetic field in the thermodynamic limit. Our prediction is testable in the present-day experimental setup for ultracold fermionic atoms in magneto-optical trap.
\end{abstract}


\maketitle    

\section{Introduction}
It is well known that the Coriolis force ($\vec{F}_C=2\bar{m}\vec{v}\times\vec{\Omega}$) acting on an electrically neutral particle in a rotating frame is analogues to the Lorentz force ($\vec{F}_L=q\vec{v}\times\vec{B}$) acting on a charged particle due to an external magnetic field in a rest frame of reference \cite{Juzeliunas}. This analogy gives rise to the artificial magnetism for a rotating harmonically trapped gas of electrically neutral particles in the rotating frame of reference \cite{Lin2}, in particular, when the angular speed of rotation approaches the angular trap frequency ($\Omega\rightarrow\omega_\perp$) \cite{Juzeliunas}. When the angular speed of rotation approaches the angular trap frequency, the trapping effect is cancelled by the centrifugal effect and the problem reduces to the cyclotron motion of a charged particle in a constant magnetic field \cite{Juzeliunas}. This analogy finds application of Landau level physics \cite{Landau2,Schweikhard,Stock}, such as the quantum Hall effect \cite{Wilkin}, in the rotating trapped quantum gases of electrically neutral atoms. However, in reality \cite{Matthews,Madison,Abo-Shaeer,Haljan,Zwierlein,Lin2}, the angular speed of rotation is always less than the angular trap frequency ($\Omega<\omega_\perp$). This is a requirement for the stability of the system in the rotating frame \cite{Stock}. The significant difference between the magnitude of the centrifugal force ($\bar{m}\Omega^2\vec{r}_\perp$) and that of the trapping force ($-\bar{m}\omega_\perp^2\vec{r}_\perp$) can drastically change the properties of artificial magnetism because the centrifugal force is analogous to a term $\bar{m}(\frac{qB}{2\bar{m}})^2\vec{r}_\perp$ which is nonlinear in artificial magnetic field \cite{Das}. Hence it is impossible to study Landau diamagnetism \cite{Landau} and de Haas-van Alphen effect \cite{DHVA} in a rotating frame for the trapped Fermi gas of neutral atoms even for $\Omega<\omega_\perp$.  

The artificial magnetism can be alternatively studied by exposing the harmonically trapped gas of neutral atoms to an artificial (or synthetic) magnetic field. An external artificial magnetic field has already been created for a neutral gas of trapped ultracold atoms \cite{Lin}. Hereafter any effect related to the artificial magnetic field will be coined as artificial effect over the existing name. Artificial quantum Hall effect has been observed in such a system of neutral atoms moving perpendicular to the artificial magnetic field \cite{LeBlanc}. Creation of artificial magnetic field for neutral atoms stimulated a lot of experimental and theoretical progress in the field \cite{Spielman,Dalibard,Jaksch,Huo,Yu,Babik}.  However, the elementary aspects of the artificial magnetism, such as artificial Landau diamagnetism \cite{Landau} and artificial de Haas-van Alphen effect \cite{DHVA}, have not been observed in trapped gas of neutral fermionic atoms \cite{Chaturvedi,Biswas3} exposed to an artificial magnetic field. A few such elementary aspects, on the other hand, were theoretically explored for both the 2-D Fermi system \cite{Farias} and the 3-D Fermi system \cite{Suzuki,Li} for low temperatures and weak \& strong fields.  However, unification of the artificial Landau diamagnetism and the artificial de Haas-van Alphen effect have not been even theoretically achieved for all temperatures and all strengths of the field. Hence we take up the discussion of artificial magnetism for a 3-D harmonically trapped Fermi gas in an artificial magnetic field. 

Pauli paramagnetism often comes into the discussion of magnetism in terms of the spin of individual particles (fermions) \cite{Pauli}. The particles we are considering though have individual spin, the spin does not couple with the artificial magnetic field. Hence we are not discussing the artificial Pauli paramagnetism. Moreover, Pauli paramagnetism can not be studied for a spin-polarized Fermi system although the polarization of the spin is required \cite{Giorgini} for trapping electrically neutral atoms in a magneto-optical trap. On the other hand, Pauli paramagnetism, Landau diamagnetism and de Haas-van Alphen effect have been unified for all temperatures and all magnitudes of the magnetic field for an untrapped ideal gas of electrons exposed to a uniform magnetic field \cite{Biswas}. We adopt and extend the method described in Ref.\cite{Biswas} for the unification of the artificial Landau diamagnetism and the artificial de Haas-van Alphen effect for all temperatures and all possible magnitudes of the artificial magnetic field.

Calculation in this article begins with the single-particle Hamiltonian of a 3-D harmonically trapped ideal Fermi gas of uncharged particles exposed to a uniform artificial magnetic field. We consider cylindrical symmetry of harmonic oscillations. We also consider that the artificial magnetic field is applied to the axis of the cylindrical symmetry. We calculate the grand free energy of the Fermi gas. We unify the artificial Landau diamagnetism and the artificial de Haas-van Alphen effect in a single framework for all temperatures and all magnitudes of the artificial magnetic field in the thermodynamic limit. Here-from we evaluate magnetic moment of the Fermi gas. We plot artificial magnetic field dependence of the magnetic moment of the Fermi system. We show the cross-over of de Haas-van Alphen oscillation and the saturation in this regard. Finally, we conclude.

\section{Grand free energy for a harmonically trapped ideal Fermi gas exposed to an artificial magnetic field}
Let us consider a 3-D harmonically trapped ideal Fermi gas of particles in thermodynamic equilibrium with a heat and particle reservoir of temperature $T$ and chemical potential $\mu$.  Let the average total number of particles in the gas be $N$. Let the mass and charge of each particle of the gas be $\bar{m}$ and $0$, respectively. We consider the (angular) trap frequency  ($\omega_\perp$) to be the same for oscillations along the $x$-axis and that along the $y$-axis. Let us now apply a uniform artificial magnetic field $\vec{B}=B\hat{k}$ to the Fermi system along the $z$-axis. Single-particle Hamiltonian of the system under the application of the artificial magnetic field becomes \cite{Lin}
\begin{eqnarray}\label{0}
H=\frac{(\vec{p}-q\vec{A})^2}{2\bar{m}}+\frac{1}{2}\bar{m}\omega_{\perp}^2r_{\perp}^2+\frac{1}{2}\bar{m}\omega_z^2z^2
\end{eqnarray}
where $\vec{p}$ is the generalized momentum of a particle, $q$ is the artificial charge of the particle which either can be set to $1$ or can be absorbed to $\vec{A}$ \cite{Stock}, $\vec{A}=-\frac{1}{2}\vec{r}\times B\hat{k}$ is the artificial gauge field, $\vec{r}=x\hat{i}+y\hat{j}+z\hat{k}$ is the position of the particle, $\vec{r}_\perp$ is the position of the particle in the $x-y$ plane, $\omega_z$ is the trap frequency of the particle along the $z$-axis. While the first term of the single-particle Hamiltonian leads to the orbital motion, the second and third terms lead to the simple harmonic oscillations.

Energy eigenvalues for the single-particle Hamiltonian of the system with respect to the lab-fixed (non-rotating) frame are given by the combination of the Fock-Darwin energy levels\footnote{Fock-Darwin energy levels are linked to $\sqrt{\omega_\perp^2+\Omega_B^2}\pm\Omega_B$ for the orbital motion and harmonic motion together in the $x-y$ plane.} and 1-D simple harmonic oscillator energy levels\footnote{Simple harmonic oscillator's energy levels are linked to $\omega_z$ for the harmonic motion along the $z$-axis.}, as \cite{Halonen}
\begin{eqnarray}\label{1}
\epsilon_{n,m,j}&=&(n+1/2)\hbar\big(\sqrt{\omega_\perp^2+\Omega_B^2}+\Omega_B\big)+(m+1/2)\nonumber\\&&\times\hbar\big(\sqrt{\omega_\perp^2+\Omega_B^2}-\Omega_B\big)+(j+1/2)\hbar\omega_z
\end{eqnarray}
where $\Omega_B=\frac{qB}{2\bar{m}}$ is half of the cyclotron frequency, $n=0,1,2,...$ represents the Landau level quantum number for $\omega_\perp\rightarrow\Omega_B$, $n-m=n,n-1,n-2,...$ represents the magnetic quantum number for $m=0,1,2,...$, and $j=0,1,2,...$ represents 1-D simple harmonic oscillator energy level quantum number. The grand free energy of the Fermi system is given by
\begin{eqnarray}\label{2}
\Phi=-k_BT\sum_{n=0}^\infty\sum_{m=0}^\infty\sum_{j=0}^\infty\ln(1+\text{e}^{-[\epsilon_{n,m,j}-\mu]/k_BT}).
\end{eqnarray}
Taylor expansion of the above logarithm about $\text{e}^{\mu/k_BT}=0$ and successive evaluation of the above summations, results in \cite{Dey}   
\begin{eqnarray}\label{3}
\Phi&=&-k_BT\bigg[\sum_{l=1}^{\infty}\frac{(-1)^{l-1}z^l}{l}\frac{1}{2\sinh(\frac{l\hbar[\sqrt{\omega_\perp^2+\Omega_B^2}+\Omega_B]}{2k_BT})}\nonumber\\&&\times\frac{1}{2\sinh(\frac{l\hbar[\sqrt{\omega_\perp^2+\Omega_B^2}-\Omega_B]}{2k_BT})}\frac{1}{2\sinh(\frac{l\hbar\omega_z}{2k_BT})}\bigg]
\end{eqnarray}
where $z=\text{e}^{\mu/k_BT}$ is the fugacity of the system. This result is exact for all temperatures and all magnitudes of the artificial magnetic field. 

The average total number of particles ($N=-\frac{\partial\Phi}{\partial\mu}|_{T,\omega_\perp,\omega_z,\Omega_B}$) in the system can be obtained from Eqn.(\ref{3}), as \cite{Dey}
\begin{eqnarray}\label{4}
N&=&\sum_{l=1}^{\infty}(-1)^{l-1}z^l\frac{1}{2\sinh(\frac{l\hbar[\sqrt{\omega_\perp^2+\Omega_B^2}+\Omega_B]}{2k_BT})}\nonumber\\&&\times\frac{1}{2\sinh(\frac{l\hbar[\sqrt{\omega_\perp^2+\Omega_B^2}-\Omega_B]}{2k_BT})}\frac{1}{2\sinh(\frac{l\hbar\omega_z}{2k_BT})}.
\end{eqnarray}
This form of the average total number of particle, however, can be simplified in the thermodynamic limit and in the limiting case of $\Omega_B\rightarrow0$, as
\begin{eqnarray}\label{5}
N\simeq\bigg(\frac{k_BT}{\hbar\bar{\omega}}\bigg)^3(-)\text{Li}_{3}(-z)
\end{eqnarray}
where $\bar{\omega}=[\omega_\perp^2\omega_z]^{1/3}$ is the geometric mean of the trap frequencies and $\text{Li}_{\nu}(z)=z+\frac{z^2}{2^\nu}+\frac{z^3}{3^\nu}+...$ is the poly-logarithmic function of the order $\nu$ and argument $z$. Here, by the thermodynamic limit, we mean $N\rightarrow\infty$, $\frac{\hbar\omega_\perp}{k_BT}\rightarrow0$, $\frac{\hbar\omega_z}{k_BT}\rightarrow0$ and $\frac{\hbar\bar{\omega}}{k_BT}N^{1/3}=constant$. The fugacity, as well as the chemical potential, is usually determined from Eqn.(\ref{5}) in terms of $T$, $\bar{\omega}$ and $N$ in the thermodynamic limit for $\Omega_B\rightarrow0$ \cite{Biswas2}. The Fermi temperature of the system can also be determined from Eqn.(\ref{5}) as $T_F=\frac{\hbar\bar{\omega}}{k_B}[6N]^{1/3}$ \cite{Biswas2}. However, extraction of the properties of the artificial magnetism needs to go beyond $\Omega_B\rightarrow0$ limit.

\section{Artificial magnetic moment of the harmonically trapped ideal Fermi gas}
\begin{figure}
\includegraphics[width=.98 \linewidth]{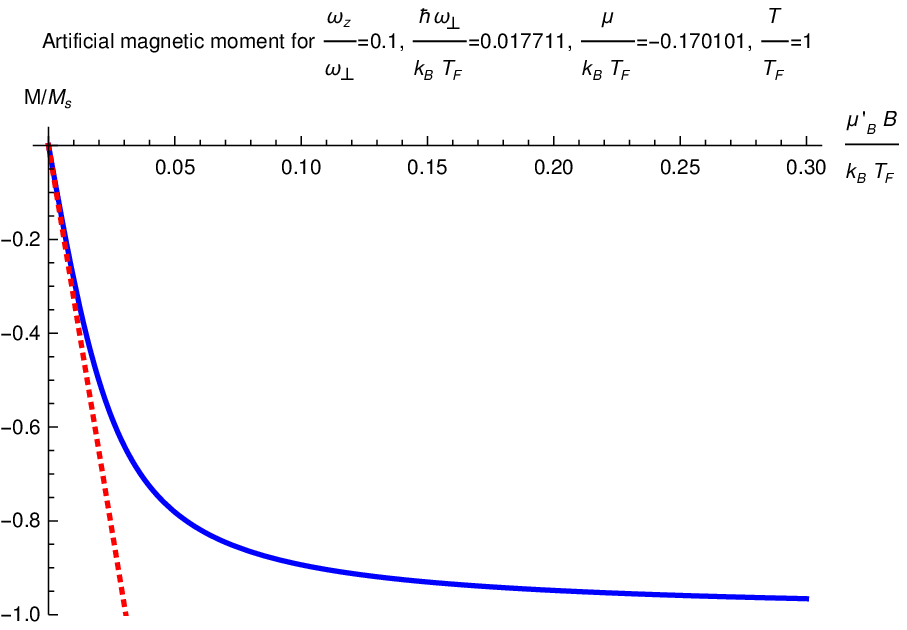}
\caption{The plot of the artificial magnetic moment versus artificial magnetic field for the harmonically trapped ideal Fermi gas for Fermi temperature $T_F=27.0976~\mu$K \cite{Stock}. While the solid line follows Eqn.(\ref{7}), the dotted line follows Eqn.(\ref{8}).
\label{fig1}}
\end{figure}

The artificial magnetic moment of the system of our interest can be defined by $\vec{M}=-\frac{\partial\Phi}{\partial\vec{B}}|_{\mu,T}=-\hat{k}\frac{q}{2\bar{m}}\frac{\partial\Phi}{\partial\Omega_B}|_{\mu,T}$. From Eqn.(\ref{3}), by following this definition, we get value of the artificial magnetic moment of the Fermi system along the $z$-direction, as
\begin{eqnarray}\label{6}
M&=&\frac{\hbar q}{2\bar{m}}\bigg[\sum_{l=1}^{\infty}\frac{(-1)^{l-1}z^l\text{csch}(\frac{l\hbar\omega_z}{2k_BT})}{4}\times\nonumber\\&&\frac{\sqrt{\omega_\perp^2+\Omega_B^2}\sinh(\frac{l\hbar\Omega_B}{k_BT})-\Omega_B\sinh(\frac{l\hbar\sqrt{\omega_\perp^2+\Omega_B^2}}{k_BT})}{\sqrt{\omega_\perp^2+\Omega_B^2}\big[\cosh(\frac{l\hbar\Omega_B}{k_BT})-\cosh(\frac{l\hbar\sqrt{\omega_\perp^2+\Omega_B^2}}{k_BT})\big]^2}\bigg].~~~~
\end{eqnarray}
This form of artificial magnetic moment is exact in all respects. This form, of course, takes the finite-size effect into account. It is expected that the artificial magnetic moment would be proportional to the average total number of particles, at least, in the thermodynamic limit. Exact form of the artificial magnetic moment per particle can now be obtained from Eqns.(\ref{6}) and  (\ref{4}), as
\begin{eqnarray}\label{7}
\frac{M}{N}&=&\frac{\hbar q}{2\bar{m}}\bigg[\sum_{l=1}^{\infty}\frac{(-1)^{l-1}z^l\text{csch}(\frac{l\hbar\omega_z}{2k_BT})}{4}\times\nonumber\\&&\frac{\sqrt{\omega_\perp^2+\Omega_B^2}\sinh(\frac{l\hbar\Omega_B}{k_BT})-\Omega_B\sinh(\frac{l\hbar\sqrt{\omega_\perp^2+\Omega_B^2}}{k_BT})}{\sqrt{\omega_\perp^2+\Omega_B^2}\big[\cosh(\frac{l\hbar\Omega_B}{k_BT})-\cosh(\frac{l\hbar\sqrt{\omega_\perp^2+\Omega_B^2}}{k_BT})\big]^2}\bigg]/\nonumber\\&&\bigg[\sum_{l=1}^{\infty}(-1)^{l-1}z^l\frac{1}{2\sinh(\frac{l\hbar[\sqrt{\omega_\perp^2+\Omega_B^2}+\Omega_B]}{2k_BT})}\nonumber\\&&\times\frac{1}{2\sinh(\frac{l\hbar[\sqrt{\omega_\perp^2+\Omega_B^2}-\Omega_B]}{2k_BT})}\frac{1}{2\sinh(\frac{l\hbar\omega_z}{2k_BT})}\bigg].
\end{eqnarray}
This equation suggests that the artificial magnetic moment of the system has a saturation value $-M_s$ where $M_s=N\frac{\hbar q}{2\bar{m}}$. The saturation value, of course, is reached at $B\rightarrow\infty$. Let us now define the artificial magneton as $\mu_B'=\frac{\hbar q}{2\bar{m}}$ which is the counter of the Bohr magneton for the artificial magnetism. The artificial magnetic moment can be expressed in terms of this quantity specially for the case of the weak artificial magnetic field ($\mu_B'B\ll k_BT$). Plotting of Eqn.(\ref{7}) with respect to the magnitude of the artificial magnetic field needs the chemical potential to be evaluated from Eqn.(\ref{4}) as a function of $N$, $T$, and the trap frequencies. We plot Eqn.(\ref{7}) (solid line) in figure \ref{fig1} for the parameters as shown in the plot-label and that mentioned in the figure caption. These parameters, in particular, $\omega_z$, $\omega_\perp$, and $T_F$ are taken from the experimental values for $N=3\times10^5$ \cite{Stock}. It is clear from figure \ref{fig1} that the artificial magnetic moment is negative and reaches the saturation as $B\rightarrow\infty$. This result is analogous to the phenomenon of orbital magnetism. For the case of the low temperature ($T\ll T_F$) and strong artificial magnetic field ($k_BT\lesssim\mu_B'B\lnsim k_BT_F$), the artificial magnetic moment would be found to be oscillating with respect to the changes in $B$ due to the existence of alternating series of the powers of $z$ ($\gg1$) in Eqn.(\ref{6}) and the competition of the two terms $\sqrt{\omega_\perp^2+\Omega_B^2}\sinh(\frac{l\hbar\Omega_B}{k_BT})$ and $\Omega_B\sinh(\frac{l\hbar\sqrt{\omega_\perp^2+\Omega_B^2}}{k_BT})$ in numerator of the same equation. This phenomenon would be coined as the artificial de Haas-van Alphen effect. 

\subsection{Artificial Landau diamagnetism}

We recast Eqn.(\ref{7}) to the lowest leading order in $\Omega_B$ after the Laurent series expanding the hyperbolic functions (e.g. $\sinh(x)=x+\frac{x^3}{3!}+\frac{x^5}{5!}+...$ and $\text{cosh}(x)=\frac{1}{x}-\frac{x}{6}+\frac{7x^3}{360}-...$), as
\begin{eqnarray}\label{8}
M\simeq-M_s\frac{-\text{Li}_{2}(-z)}{-\text{Li}_{3}(-z)}\frac{\hbar\Omega_B}{3k_BT}=-M_s\frac{-\text{Li}_{2}(-z)}{-\text{Li}_{3}(-z)}\frac{\mu_B'B}{3k_BT}.
\end{eqnarray}
This result corresponds to Landau diamagnetism. Temperature dependence of the fugacity of Eqn.(\ref{8}) can be determined from Eqn.(\ref{5}). There-from it can be shown that $\frac{-\text{Li}_{2}(-z)}{-\text{Li}_{3}(-z)}\rightarrow1$ for $T/T_F\rightarrow\infty$. Hence  Eqn.(\ref{8}) follows Curie's law at the high temperatures. For  $T/T_F\rightarrow0$, on the hand, we have $\frac{-\text{Li}_{2}(-z)}{-\text{Li}_{3}(-z)}\rightarrow\frac{3T}{T_F}$ according to the Sommerfeld's asymptotic expansions of both the numerator (Fermi-Dirac integral of order $2$) and the denominator (Fermi-Dirac integral of order $3$) \cite{Biswas2}. Thus for $T/T_F\rightarrow0$, the artificial magnetic moment takes the form $M\simeq-M_s\frac{\mu_B'B}{k_BT_F}$ to the lowest leading order in $B$. We plot Eqn.(\ref{8}) (dotted line) in figure \ref{fig1} for the parameters same as that for the solid line in the same figure. Linearity of the plot with $B$ and negative values of $M$ confirm the existence of the artificial Landau diamagnetism for the harmonically trapped ideal Fermi gas of uncharged particles exposed to an artificial magnetic field. This result would, however, be okay for the case of the weak field ($\frac{\mu_B'B}{k_BT}\ll1$). The solid line in the same figure, on the other hand, captures the finite-size effect and the higher order effect of $B$, and consequently, exhibits the saturation of the artificial magnetic moment of the system. 

\subsection{Artificial de Haas-van Alphen effect}
Artificial magnetic field dependence of the artificial magnetic moment would be significantly different at low temperature and the strong artificial magnetic field regime ($k_BT\lesssim\mu_B'B\lnsim k_BT_F$). Artificial de Haas-van Alphen oscillation would be observed in this regime. We will, however, investigate the same in the thermodynamic limit. Properties of the artificial de Haas-van Alphen effect is supposed to be extracted from Eqn.(\ref{6}) because it represents an exact result even for low temperatures and strong fields. Although Eqn.(\ref{6}) is plausible for high temperature and weak field regime ($\mu_B'B\ll k_BT$), it is not plausible for low temperatures due to the existence of a pole type singularity in $M/N$ at either $T\rightarrow0$ or at $\Omega_B\rightarrow\infty$. We can avoid such a problem by recasting the grand free energy in Eqn.(\ref{3}) for low temperatures and strong artificial magnetic fields.

Since we have $\frac{1}{2\sinh(x/2)}=\frac{\text{e}^{-x/2}}{1-\text{e}^{-x}}$ $\forall~x\in\mathbb{R}$, we recast Eqn.(\ref{3}), as
\begin{eqnarray}\label{9}
\Phi&=&-k_BT\bigg[\sum_{l=1}^{\infty}\frac{(-1)^{l-1}z^l}{l}\frac{\text{e}^{-\frac{l\hbar[\sqrt{\omega_\perp^2+\Omega_B^2}+\Omega_B]}{2k_BT}}}{1-\text{e}^{-\frac{l\hbar[\sqrt{\omega_\perp^2+\Omega_B^2}+\Omega_B]}{k_BT}}}\nonumber\\&&\times\frac{\text{e}^{-\frac{l\hbar[\sqrt{\omega_\perp^2+\Omega_B^2}-\Omega_B]}{2k_BT}}}{1-\text{e}^{-\frac{l\hbar[\sqrt{\omega_\perp^2+\Omega_B^2}-\Omega_B]}{k_BT}}}\frac{\text{e}^{-\frac{l\hbar\omega_z}{2k_BT}}}{1-\text{e}^{-\frac{l\hbar\omega_z}{k_BT}}}\bigg].
\end{eqnarray}
The above equation can be further recast by truncating the Taylor expansions of the exponentials in the last two fractions and taking the binomial expansion of $\frac{1}{1-\text{e}^{-\frac{l\hbar[\sqrt{\omega_\perp^2+\Omega_B^2}+\Omega_B]}{k_BT}}}$, as
\begin{eqnarray}\label{10}
\Phi&\simeq&-k_BT\bigg[\sum_{l=1}^{\infty}\frac{(-1)^{l-1}z^l}{l}\frac{1}{\frac{l\hbar[\sqrt{\omega_\perp^2+\Omega_B^2}-\Omega_B]}{k_BT}}\frac{1}{\frac{l\hbar\omega_z}{k_BT}}\nonumber\\&&\times\text{e}^{-\frac{l\hbar[\sqrt{\omega_\perp^2+\Omega_B^2}+\Omega_B]}{2k_BT}}\sum_{k=1}^{\infty}\text{e}^{-\frac{l(k-1)\hbar[\sqrt{\omega_\perp^2+\Omega_B^2}+\Omega_B]}{2k_BT}}\bigg]\nonumber\\&\simeq&-\frac{(k_BT)^3[\sqrt{\omega_\perp^2+\Omega_B^2}+\Omega_B]}{\hbar^2\omega_\perp^2\omega_z}\nonumber\\&&\times\sum_{k=1}^{\infty}(-)\text{Li}_{3}\big(-\text{e}^{\frac{\mu-(k-1)\hbar\Omega_B-k\hbar\sqrt{\omega_\perp^2+\Omega_B^2}}{k_BT}}\big)
\end{eqnarray}
in the thermodynamic limit. Here-from we get the average total number of particles  ($N=-\frac{\partial\Phi}{\partial\mu}|_{T,\omega_\perp,\omega_z,\Omega_B}$), as 
\begin{eqnarray}\label{11}
N&\simeq&\frac{(k_BT)^2[\sqrt{\omega_\perp^2+\Omega_B^2}+\Omega_B]}{\hbar^2\omega_\perp^2\omega_z}\nonumber\\&&\times\sum_{k=1}^{\infty}(-)\text{Li}_{2}\big(-\text{e}^{\frac{\mu-k\hbar\sqrt{\omega_\perp^2+\Omega_B^2}-(k-1)\hbar\Omega_B}{k_BT}}\big).
\end{eqnarray}
Exact expression of magnitude of the artificial magnetic moment ($\vec{M}=-\frac{\partial\Phi}{\partial\vec{B}}|_{\mu,T}=-\hat{k}\frac{q}{2\bar{m}}\frac{\partial\Phi}{\partial\Omega_B}|_{\mu,T}$) can be obtained from Eqn.(\ref{9}), as
\begin{eqnarray}\label{12}
M&=&\sum_{l=1}^\infty\frac{\mu_B'(-1)^{l}\text{e}^{\frac{\mu l-l\hbar\sqrt{\omega_\perp^2+\Omega_B^2}}{k_BT}}}{\big[1-\text{e}^{\frac{-l\hbar[\sqrt{\omega_\perp^2+\Omega_B^2}-\Omega_B]}{k_BT}}\big]^2\big[1-\text{e}^{\frac{-l\hbar[\sqrt{\omega_\perp^2+\Omega_B^2}+\Omega_B]}{k_BT}}\big]^2}\nonumber\\&&\times\frac{\text{e}^{-\frac{l\hbar\omega_z}{2k_BT}}}{1-\text{e}^{-\frac{l\hbar\omega_z}{k_BT}}}\bigg(\frac{\Omega_B}{\sqrt{\omega_\perp^2+\Omega_B^2}}\bigg[1-\text{e}^{-\frac{2l\hbar\sqrt{\omega_\perp^2+\Omega_B^2}}{k_BT}}\bigg]\nonumber\\&&-\bigg[\text{e}^{-\frac{l\hbar[\sqrt{\omega_\perp^2+\Omega_B^2}-\Omega_B]}{k_BT}}-\text{e}^{-\frac{l\hbar[\sqrt{\omega_\perp^2+\Omega_B^2}+\Omega_B]}{k_BT}}\bigg]\bigg).
\end{eqnarray}
It is clear from the above equation that the artificial magnetic moment of the system is always negative. It is also clear from the terms in the parenthesis of the above equation that the artificial magnetic moment would be linear in $\Omega_B$ for small $\Omega_B$s as expected from the artificial Landau diamagnetism.

\begin{figure}
\includegraphics[width=.98 \linewidth]{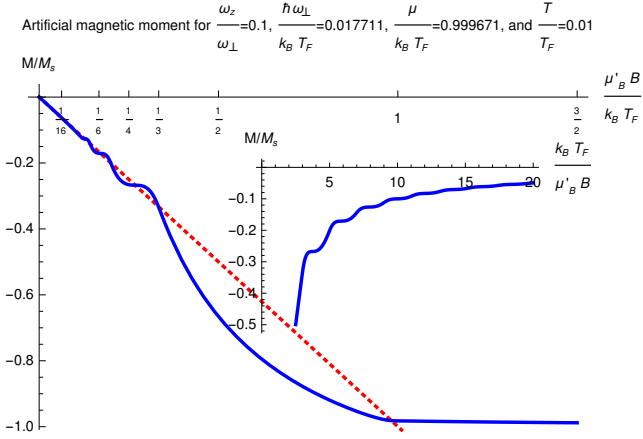}
\caption{The plot of the artificial magnetic moment versus artificial magnetic field for the harmonically trapped ideal Fermi gas for Fermi temperature $T_F=27.0976~\mu$K \cite{Stock}. The solid line follows Eqn.(\ref{13}). The dotted line follows Eqn.(\ref{8}). The solid line in the inset also follows Eqn.(\ref{13}).
\label{fig2}}
\end{figure}

Now approximating $\frac{\text{e}^{-\frac{l\hbar\omega_z}{2k_BT}}}{1-\text{e}^{-\frac{l\hbar\omega_z}{k_BT}}}$ as $\frac{1}{\frac{l\hbar\omega_z}{k_BT}}$, $\frac{1}{\big[1-\text{e}^{-\frac{l\hbar[\sqrt{\omega_\perp^2+\Omega_B^2}-\Omega_B]}{k_BT}}\big]^2}$ as $\frac{1}{\big[\frac{l\hbar[\sqrt{\omega_\perp^2+\Omega_B^2}-\Omega_B]}{k_BT}\big]^2}$, binomial expanding $\frac{1}{\big[1-\text{e}^{-\frac{l\hbar[\sqrt{\omega_\perp^2+\Omega_B^2}+\Omega_B]}{k_BT}}\big]^2}$ as $\sum_{k=1}^\infty k\text{e}^{-\frac{(k-1)l\hbar[\sqrt{\omega_\perp^2+\Omega_B^2}+\Omega_B]}{k_BT}}$, and evaluating the summation over $l$, we recast Eqn.(\ref{12}) with the further use of Eqn.(\ref{11}), as
\begin{eqnarray}\label{13}
\frac{M}{N}&\simeq&-\mu_B'\frac{k_BT}{\hbar[\sqrt{\omega_\perp^2+\Omega_B^2}-\Omega_B]}\bigg\{\sum_{k=1}^\infty k\bigg(\frac{\Omega_B}{\sqrt{\omega_\perp^2+\Omega_B^2}}\nonumber\\&&\times\bigg[(-)\text{Li}_3\big(-\text{e}^{\frac{\mu-\hbar[k\sqrt{\omega_\perp^2+\Omega_B^2}+(k-1)\Omega_B]}{k_BT}}\big)\nonumber\\&&-(-)\text{Li}_3\big(-\text{e}^{\frac{\mu-\hbar[(k+2)\sqrt{\omega_\perp^2+\Omega_B^2}+(k-1)\Omega_B]}{k_BT}}\big)\bigg]\nonumber\\&&+\bigg[(-)\text{Li}_3\big(-\text{e}^{\frac{\mu-\hbar[(k+1)\sqrt{\omega_\perp^2+\Omega_B^2}+k\Omega_B]}{k_BT}}\big)\nonumber\\&&-(-)\text{Li}_3\big(-\text{e}^{\frac{\mu-\hbar[(k+1)\sqrt{\omega_\perp^2+\Omega_B^2}+(k-2)\Omega_B]}{k_BT}}\big)\bigg]\bigg)\bigg\}/\nonumber\\&&\bigg[\sum_{n=1}^{\infty}(-)\text{Li}_{2}\big(-\text{e}^{\frac{\mu-\hbar[n\sqrt{\omega_\perp^2+\Omega_B^2}+(n-1)\hbar\Omega_B]}{k_BT}}\big)\bigg]
\end{eqnarray}
in the thermodynamic limit. Eqn.(\ref{13}) represents the artificial magnetic moment per particle of the Fermi system. This equation is useful for describing the artificial de Haas-van Alphen effect for strong artificial magnetic fields. This equation is also useful for describing the artificial magnetism for weak artificial magnetic fields in the thermodynamic limit. Eqn.(\ref{11}) becomes same as Eqn.(\ref{5}) once the summation over $k$ is replaced by integration over $k$ in the thermodynamic limit for $\Omega_B\rightarrow0$. Now we evaluate the chemical potential from Eqn.(\ref{5}) according to the method described in Ref.\cite{Biswas2}. We plot the artificial magnetic moment of Eqn.(\ref{13}) in figure \ref{fig2} (solid line) with respect to the artificial magnetic field for the parameters same as that used in figure \ref{fig1} except the temperature $T=0.01~T_F$ and chemical potential $\mu=0.999671~k_BT_F$. We also plot the artificial magnetic moment of Eqn.(\ref{8}) (dotted) line in the same figure for the same parameters as mentioned in the plot-label and the figure caption. We also plot the artificial magnetic moment with respect to inverse of the artificial magnetic field in the inset of figure \ref{fig2} for same set of parameters. Oscillations in the artificial magnetic moment about its weak field results, which are indicated by the dotted line, represent the artificial de Haas-van Alphen effect.  While de Haas-van Alphen oscillations are observed to be periodic with the inverse of the external magnetic field in a gas of free electrons in a metal \cite{DHVA}, the artificial de Haas-van Alphen oscillations are not appearing periodic with the inverse of the artificial magnetic field, rather quasi-periodic due to the inhomogeneity of the Fermi system, as clear from the inset of figure \ref{fig2}. Both the artificial Landau diamagnetism and the artificial de Haas-van Alphen effect are described by the solid line in figure \ref{fig2}. Hence  Eqn.(\ref{13}) unifies the  artificial Landau diamagnetism and the artificial de Haas-van Alphen effect in the thermodynamic limit.

\begin{figure}
\includegraphics[width=.90 \linewidth]{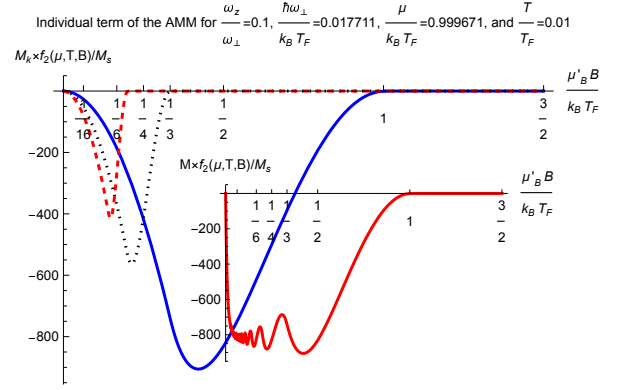}
\caption{The plot of the $k$th term ($M_k$) of the artificial magnetic moment (AMM) versus the  artificial magnetic field for $k=1$ (solid line), $2$ (dotted line), and $3$ (dashed line). The $k$th term $M_k$ comes from Eqn.(\ref{13}) for the parameters as mentioned in figure \ref{2}. The inset represents the sum of all the $M_k$ terms ($M=\sum_{k=1}^{\infty}M_k$) and the summation follows Eqn.(\ref{13}) for the same parameters.
\label{fig3}}
\end{figure} 

What makes the artificial de Haas-van Alphen oscillation for the trapped system of our interest is not very clear from Eqn.(\ref{13}). We can analyse the oscillation by focusing on the individual terms ($\{M_kf_{2}(\mu,T,B)/M_s\}$, $k=1,2,3,...$) of the artificial magnetic moment ($M$) in units of $M_s/f_{2}(\mu,T,B)$ where $f_{2}(\mu,T,B)=\sum_{n=1}^{\infty}(-)\text{Li}_{2}\big(-\text{e}^{\frac{\mu-\hbar[n\sqrt{\omega_\perp^2+\Omega_B^2}+(n-1)\hbar\Omega_B]}{k_BT}}\big)$ is the non-oscillating denominator of $M$ as expressed in Eqn.(\ref{13}). We plot $M_k$ in figure \ref{fig3} for $k=1$ (solid line), $2$ (dotted line) and, $3$ (dashed line) with respect to the artificial magnetic field. The polylogarithmic functions (Fermi integrals $\{-\text{Li}_3\}$) in $M_k$ though individually diverge very slowly (cubic power of logarithm \cite{Biswas2}) for $T\rightarrow0$ and $B\rightarrow0$, the operations on the polylogarithmic functions result in $M_k$ to start decreasing at $B=0$ and increasing at around $\frac{\mu_B'B}{k_BT_F}/k\lnsim1$ (strong field regime) before slowly vanishing beyond $\frac{\mu_B'B}{k_BT_F}/k\simeq1$ at a low temperature ($T/T_F\ll1$, $\mu\simeq k_BT_F$) for $\hbar\omega_\perp\ll k_BT\lesssim \mu_B'B$. This leads to an oscillation of the artificial magnetic moment ($M=\sum_{k=1}^{\infty}M_k$) with a quasi-period inversely proportional to the artificial magnetic field in the interval $(0,k_BT_F/\mu_B')$, as clear from the inset of figure \ref{fig3}.

\section{Conclusion}
To conclude, we have analytically explored the artificial magnetism for a 3-D spin-polarized harmonically trapped ideal Fermi gas of electrically neutral particles exposed to a uniform synthetic magnetic field. We have unified artificial Landau diamagnetism and artificial de Haas-van Alphen effect for such a system. Our unification is applicable for all temperatures as well as for all strengths of the synthetic magnetic field in the thermodynamic limit. Our prediction is testable within the existing experimental setup for ultracold fermionic atoms in magneto-optical trap \cite{Stock,Lin,Lin2}.

We could study artificial magnetism rotating harmonically trapped Bose gas of uncharged particles. All the results would be unaltered for this case except the replacement of $(-1)^lz^l$ by $z^l$ and the replacement of $-\text{Li}_j(-x)$ by $\text{Li}_j(x)$ $\forall~x\in~\mathbb{R}$. However, necessary condition for observing the artificial de Haas-van Alphen effect would be $\mu>\epsilon_{0,0,0}$. Such a condition is not reached by a Bose system. Hence one can not observe artificial de Haas-van Alphen effect in a Bose system.

Properties of rotating trapped Bose and Fermi systems are often analysed with respect to the body-fixed rotating frame of reference \cite{Stock,Das}. We have, however, analysed the artificial magnetic properties of the Fermi system from the lab-fixed (non-rotating) frame. The artificial magnetism would have been studied with respect to the body-fixed rotating frame without applying the artificial magnetic field \cite{Das}. However, the centrifugal force additionally appears in the rotating frame and consequently, it changes the entire physics because of appearance of a pole type singularity at $\Omega_B\rightarrow\omega_\perp$ in the grand free energy of the system. It also brings the question of stability of the system in the rotating frame for $\Omega_B\ge\omega_\perp$. We would not get the artificial Landau diamagnetism, as well as Curie's law, in the rotating frame. 

We have considered the grand canonical ensemble for obtaining our results. Consequently, we have fixed the chemical potential ($\mu$) for our analyses. However, the average number of particles ($N$) becomes fixed once we fix the chemical potential and the temperature ($T$) of the system.   

We did not directly consider any accidental degeneracy because the single-particle energy eigenvalues as mentioned in Eqn.(\ref{1}) are non-degenerate due to the trapping effect for a finite strength of the artificial magnetic field. The accidental degeneracy significantly modifies the density of states and consequently plays a key role in the de Haas-van Alphen effect on an untrapped ideal gas of electrons \cite{Landau,Biswas}. However, a similar accidental degeneracy is also there with quantum number $m$ (Eqn.(\ref{1})) of the trapped system of our interest in the limiting case of $\frac{\omega_\perp}{\Omega_B}\rightarrow0$ for a strong artificial magnetic field.

We have not considered the inter-particle interaction in our analysis. While the weak attractive inter-particle interactions lead to the formation of the Cooper pairs in the harmonically trapped Fermi gas \cite{Giorgini,Parish}, the strong attractive interactions result in the BEC-BCS crossover in the same system \cite{Parish,Randeria}. The study of the artificial magnetism for the harmonically trapped interacting Fermi gas of uncharged particles exposed to an artificial magnetic field is kept as an open problem.

Pauli paramagnetism, Landau diamagnetism, and de Haas-van Alphen effect have already been unified for an untrapped ideal gas of electrons \cite{Biswas}. Our consideration of only one component of spin of the uncharged fermionic atoms stopped us from studying Pauli paramagnetism. One component of the spin is required for trapping the atoms in a magneto-optical trap \cite{Pitaevskii,Giorgini}. However, how to unify Pauli paramagnetism, Landau diamagnetism, and de Haas-van Alphen effect for the artificial magnetism of a multi-spin component trapped Fermi gas, is kept as an open problem.

\acknowledgments
S. Biswas acknowledges partial financial support of the SERB, DST, Govt. of India under the EMEQ Scheme [No. EEQ/2019/000017]. Useful discussions with Prof. J. K. Bhattacharjee (IACS, Kolkata) and Mr. Samir Das (UoH, Hyderabad) are gratefully acknowledged. We thank the anonymous reviewers for their thorough review and highly appreciate their comments and suggestions which significantly contributed to improving the quality of the presentation.

\end{document}